\documentclass[12pt]{iopart}
\usepackage{graphicx}
\usepackage{url}

\def\dx{\mathrm{d}x}
\def\inner#1#2{ \langle #1, #2 \rangle }
\def\Inner#1#2{ \left\langle #1, #2 \right\rangle }
\def\E{\mathcal{E}}
 
\begin{document}

\title{A closed-form expression for the Sharma-Mittal entropy of exponential families}
\author{Frank Nielsen$^{1,2}$ and Richard Nock$^3$}

\address{$^1$ Sony Computer Science Laboratories, Inc. Tokyo, Japan.\\ $^2$ \'Ecole Polytechnique, CS Dept. LIX, Palaiseau, France.}

\address{$^3$ UAG-CEREGMIA, Martinique, France.}

\eads{\mailto{Frank.Nielsen@acm.org}, \mailto{Richard.Nock@martinique.univ-ag.fr}}

\begin{abstract}
The Sharma-Mittal entropies generalize the celebrated Shannon, R\'enyi and Tsallis entropies.
We report a closed-form formula for the Sharma-Mittal entropies and relative entropies for arbitrary exponential family distributions.
We instantiate explicitly the formula for the case of the multivariate Gaussian distributions and discuss on its estimation.
\end{abstract}

\submitto{Journal of Physics A: Mathematical and Theoretical\\Fast Track Communications  (revised manuscript, Nov. 2011; accepted Nov. 2011)}

\maketitle

The Sharma-Mittal entropy $H_{\alpha,\beta}(p)$~\cite{SharmaMittal-1975,SharmaMittal-2007} of a probability density\footnote{For sake of simplicity and without loss of generality, we consider the probability density function $p$ of a continuous random variable $X\sim p$ in this note.
For multivariate densities $p$, the integral notation $\int$ denote the corresponding multi-dimensional integral, so that we write for short $\int p(x)\dx =1$. 
Our results hold for probability mass functions, and probability measures in general.} $p$  is defined as  

\begin{equation}\label{eq:defsm}
H_{\alpha,\beta}(p) = \frac{1}{1-\beta} \left( \left(\int p(x)^\alpha \dx\right)^{\frac{1-\beta}{1-\alpha}} -1 \right),\mbox{with\ } \alpha>0, 
\alpha\not =1, \beta\not=1.
\end{equation}

This bi-parametric family of  entropies tends in limit cases to 
R\'enyi entropies $R_\alpha(p)=\frac{1}{1-\alpha} \log \int p(x)^{\alpha}\dx$ (for $\beta\rightarrow 1$),  Tsallis entropies
$T_\alpha(p)=\frac{1}{1-\alpha}(\int p(x)^{\alpha}\dx -1)$ (for $\beta\rightarrow\alpha$),
and  Shannon entropy $H(p)=-\int p(x)\log p(x) \dx$ (for both $\alpha,\beta\rightarrow 1$).
Sharma-Mittal entropy has previously been studied in  
the context
of multidimensional harmonic oscillator systems~\cite{SobolevSharmaMittal:2008}.

Many usual statistical distributions including the Gaussians and discrete multinomials (that is, normalized histograms)  belong to the exponential families~\cite{Brown:1986}. 
Those exponential families play a major role in the field of thermo-statistics~\cite{Naudts-2011}, and admit the  generic canonical decomposition

\begin{equation}\label{eq:expfam}
p_F(x | \theta) = \exp \left( \inner{\theta}{t(x)}-F(\theta)+k(x) \right),
\end{equation}
where $\inner{\cdot}{\cdot}$ denote the inner product, $F$ is a strictly convex $C^\infty$ function characterizing the family (called the log-normalizer since $F(\theta)=\log \int e^{\inner{\theta}{t(x)}+k(x)} \dx$), $\theta\in\Theta$ is the natural parameter denoting the member of the family $\E_F=\{p_F(x|\theta)\ |\ \theta\in\Theta \}$, $t(x)$ is the sufficient statistics, and $k(x)$ is an auxiliary carrier measure~\cite{Brown:1986}.
The natural parameter space $\Theta=\{\theta\ |\ p_F(x;\theta)<\infty \}$ is an open convex set.

For example, the probability density of a multivariate Gaussian $p\sim N(\mu,\Sigma)$ centered at $\mu$ with positive-definite covariance matrix $\Sigma$ is conventionally  written as

\begin{equation}\label{eq:gaussian}
p(x | \mu,\Sigma) = \frac{1}{(2\pi)^{\frac{d}{2}}\sqrt{|\Sigma|}} \exp -\frac{(x-\mu)^T \Sigma^{-1}  (x-\mu) }{2},
\end{equation}
where $|\Sigma|>0$ denote the determinant of the positive definite matrix.
Rewriting  the density of Eq.~\ref{eq:gaussian} to fit the canonical decomposition of Eq.~\ref{eq:expfam}, we get

\begin{equation}
 p(x|\mu,\Sigma) = \exp \left(
-\frac{1}{2} x^T \Sigma^{-1} x + x^T \Sigma^{-1}\mu - \frac{1}{2}\mu^T\Sigma^{-1}\mu - \frac{1}{2}\log (2\pi)^{d} |\Sigma| 
\right).
\end{equation}
Using the matrix trace cyclic property, we have 
$-\frac{1}{2}x^T\Sigma^{-1}x=\tr (-\frac{1}{2}x^T\Sigma^{-1}x)=\tr(x x^T \times (-\frac{1}{2}\Sigma^{-1}))$, where $\mathrm{tr}$ denote the matrix trace operator.
It follows that

\begin{eqnarray}
 p(x|\mu,\Sigma) & = &  \exp \left( \Inner{(x,x x^T)}{\left(\Sigma^{-1}\mu,-\frac{1}{2}\Sigma^{-1}\right)} - F(\theta)\right),\\
 & = & p(x| \theta).
\end{eqnarray}
with $\theta=(\Sigma^{-1}\mu,-\frac{1}{2}\Sigma^{-1})$ and $F(\theta)=\frac{1}{2}\log (2\pi)^d|\Sigma| + \frac{1}{2}\mu^T \Sigma^{-1} \mu$  (and $k(x)=0$).
In this decomposition, the natural parameter 
$\theta=(\Sigma^{-1}\mu, -\frac{1}{2}\Sigma^{-1})=(v,M)$ consists of two parts:
 a vectorial part $v$, and a symmetric negative definite matrix part $M\prec 0$.
The inner product of $\theta=(v,M)$  and $\theta'=(v',M')$  is defined as 
$\inner{\theta}{\theta'} =  v^T v' + \mathrm{tr}(M^T M')$.
For univariate normal distributions, the natural parameter $\theta$ is $(\frac{\mu}{\sigma^2},-\frac{1}{2\sigma^2})$.
The order of the exponential family is the dimension of its natural parameter space $\Theta$.
Normal $d$-dimensional distributions $N(\mu,\Sigma)$ form an exponential family of order $d+\frac{d(d+1)}{2} = \frac{d(d+3)}{2}$.

We have $M=-\frac{1}{2}\Sigma^{-1}$, that is $|\Sigma^{-1}|=|\Sigma|^{-1}=|-2M|$, and $\mu^T=-\frac{1}{2}v^T M^{-1}$ 
(since $M^{-1}=-2\Sigma$, $-\frac{1}{2}M^{-1}v=\Sigma v = \mu$ and $M^{-T}=M^{-1}$).
It follows that the log-normalizer $F$ expressed using the canonical natural parameters  is 

\begin{eqnarray}
 F(\mu,\Sigma) & = & \frac{1}{2}\log (2\pi)^d|\Sigma| + \frac{1}{2}\mu^T \Sigma^{-1} \mu,  \label{eq:can}\\
 F(v,M) &=& \frac{d}{2}\log 2\pi - \frac{1}{2}\log |-2M|  - \frac{1}{4} v^T M^{-1} v.  \label{eq:GaussianF}
\end{eqnarray}

In order to calculate the Sharma-Mittal entropy of Eq.~\ref{eq:defsm}, let $M_\alpha(p)=\int p(x)^\alpha \dx$ so that

\begin{equation}
H_{\alpha,\beta}(p) = 
\frac{1}{1-\beta} \left( M_\alpha(p)^{\frac{1-\beta}{1-\alpha}} -1 \right).
\end{equation}

Let us prove that for an arbitrary exponential family $\E_F=\{p_F(x | \theta)\ |\ \theta\in\Theta \}$

\begin{equation}\label{eq:m}
M_\alpha(p)=e^{F(\alpha\theta)-\alpha F(\theta)} E_p[e^{(\alpha-1)k(x)}]
\end{equation}

\noindent Proof:
\begin{eqnarray}
M_\alpha(p) &=& \int e^{\alpha (\inner{t(x)}{\theta}-F(\theta)+k(x)) } \dx,  \\
& = & \int e^{\inner{t(x)}{\alpha\theta}-\alpha F(\theta)+\alpha k(x) + (1-\alpha) k(x) - (1-\alpha) k(x) + F(\alpha\theta)-F(\alpha\theta)} \dx, \\
& = & \int e^{F(\alpha\theta)-\alpha F(\theta)} p_F(x;\alpha\theta) e^{(\alpha-1) k(x)} \dx, \label{eq:obs1}\\
& = & e^{F(\alpha\theta)-\alpha F(\theta)}  \int p_F(x;\alpha\theta) e^{(\alpha-1) k(x)} \dx, \\
& =& e^{F(\alpha\theta)-\alpha F(\theta)}  E_p[e^{(\alpha-1) k(x)}].
\end{eqnarray}
Observe that in Eq.~\ref{eq:obs1}, we require $\alpha\theta\in\Theta$ for a valid exponential family distribution. 
This is the case whenever the natural parameter space $\Theta$ is a convex cone (e.g., Gaussian case).
It follows from Eq.~\ref{eq:m} that the Sharma-Mittal entropy of a distribution $p\sim \E_F$ belonging to an exponential family $\E_F$ is

\begin{equation} \label{eq:SMH}
H_{\alpha,\beta}(p) = \frac{1}{1-\beta} \left( e^{\frac{1-\beta}{1-\alpha} (F(\alpha\theta)-\alpha F(\theta))} (E_p[e^{(\alpha-1)k(x)}])^{\frac{1-\beta}{1-\alpha}}    -1 \right).
\end{equation}

In particular, when the auxiliary carrier measure $k(x)=0$~\cite{Brown:1986} (including the above-mentioned multivariate Gaussian family), Eq.~\ref{eq:SMH} becomes a  closed-form formula since $E_p[e^{(\alpha-1) k(x)}]=E_p[1]=1$:

\begin{eqnarray}  
H_{\alpha,\beta}(p) & =& \frac{1}{1-\beta} \left( e^{(1-\beta)\frac{F(\alpha\theta)-\alpha F(\theta))}{1-\alpha}}  -1 \right), \label{eq:SMEnok}\\
& = & \frac{1}{1-\beta} \left(
 (e^{F(\alpha\theta)-\alpha F(\theta)})^{\frac{1-\beta}{1-\alpha}} -1
\right).
\end{eqnarray}

We derive in limit cases expressions for the R\'enyi, Tsallis and Shannon entropies of an arbitrary  exponential family (with $k(x)=0$):
\begin{eqnarray}
R_\alpha(p) & = \lim_{\beta\rightarrow 1} H_{\alpha,\beta}(p) &=  \frac{1}{1-\alpha} (F(\alpha\theta)-\alpha F(\theta)),\\
T_\alpha(p) & = \lim_{\beta\rightarrow\alpha} H_{\alpha,\beta}(p) &=  \frac{1}{1-\alpha} (e^{F(\alpha\theta)-\alpha F(\theta)} -1),\\
H(p) & = \lim_{\beta,\alpha\rightarrow 1} H_{\alpha,\beta}(p) &=  F(\theta) - \inner{\theta}{\nabla F(\theta)}.\label{eq:HEF}
\end{eqnarray}

Note that the Shannon entropy of a member of an exponential family $p\sim \E_F$ indexed with natural parameter $\theta$ can also be rewritten as $H(p)=H(\theta)=-F^*(\eta)$ with $\eta=\nabla F(\theta)$ the dual moment coordinates, and $F^*$ the Legendre $C^\infty$ convex conjugate of $F$~\cite{informationgeometry-2000}.

\begin{figure}
\centering
\includegraphics[width=0.7\textwidth]{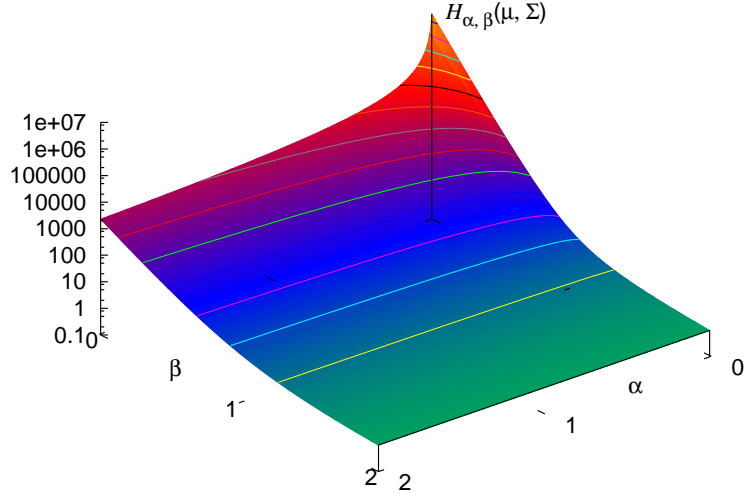}
\caption{
Plot of the Sharma-Mittal entropy (Eq.~\ref{eq:entGau}) for the $4\times 4$ covariance matrix $\Sigma=4 I$ (independent of the mean $\mu$), where $I$ denotes the identity matrix.
\label{fig:SM}}
\end{figure}

Let us instantiate the generic formula of Eq.~\ref{eq:SMEnok} to the case of multivariate Gaussians with mean parameter $\mu$ and covariance matrix $\Sigma$. We  get

\begin{equation}\label{eq:entGau}
H_{\alpha,\beta}(\mu,\Sigma) =  \frac{1}{1-\beta} \left(
\frac{ \left( (2\pi)^{\frac{d}{2}} |\Sigma|^{\frac{1}{2}}  \right)^{1-\beta} }{\alpha^{\frac{d(1-\beta)}{2(1-\alpha)}}} -1
\right),
\end{equation}
independent of $\mu$  (in 1D, $\Sigma=\sigma^2$  so that $|\Sigma|^{\frac{1}{2}}=\sigma$).
Indeed, consider the expression $F(\alpha\theta)-\alpha F(\theta)$ in Eq.~\ref{eq:SMH} for the Gaussian log-normalizer of Eq.~\ref{eq:GaussianF}.
Using the fact that $|\alpha A|=\alpha^d |A|$ for $d$-dimensional matrices $A$, the term $F(\alpha\theta)$ is

\begin{eqnarray}
F(\alpha\theta) & = &  F(\alpha v,\alpha M),\\
& = & \frac{d}{2}\log 2\pi - \frac{1}{2}\log\alpha^d|-2M|-\frac{1}{4}(\alpha v)^T(\alpha M)^{-1}(\alpha v),\\
& = &  \frac{d}{2}\log 2\pi -\frac{d}{2}\log \alpha - \frac{1}{2}\log |-2M|   -\frac{\alpha}{4} v^T M^{-1} v.\label{eq1}
\end{eqnarray}

Similarly, we have
\begin{equation}\label{eq2}
\alpha F(\theta) = \alpha F(  v,  M)= 
\frac{d\alpha}{2}\log 2\pi - \frac{\alpha}{2}\log |-2M|  - \frac{\alpha}{4} v^T M^{-1} v.
\end{equation}

Thus by subtracting Eq.~\ref{eq2} to Eq.~\ref{eq1}, we obtain

\begin{equation}
F(\alpha\theta)-\alpha F(\theta)=
\frac{d(1-\alpha)}{2}\log 2\pi - \frac{d}{2}\log\alpha-\frac{1-\alpha}{2}\log|-2M|.
\end{equation}

Therefore, we deduce that

\begin{eqnarray}
F(\alpha\theta)-\alpha F(\theta) &=& \log\left( \frac{(2\pi)^{d\frac{1-\alpha}{2}}}{\alpha^{\frac{d}{2}}} |-2M|^{-\frac{1-\alpha}{2}} \right),\\
& =  & \log \left( \frac{(2\pi)^{\frac{d}{2}(1-\alpha)}}{\alpha^{\frac{d}{2}}}  |\Sigma|^{\frac{1-\alpha}{2}} \right),
\end{eqnarray}
hence the result of Eq.~\ref{eq:entGau}.
For 1D Gaussians with standard deviation $\sigma>0$, this yields
\begin{equation}\label{eq:entGau1D}
H_{\alpha,\beta}(\mu,\sigma) =  \frac{1}{1-\beta} \left(
\frac{ \left( \sqrt{2\pi}  \sigma  \right)^{1-\beta} }{\alpha^{\frac{(1-\beta)}{2(1-\alpha)}}} -1
\right).
\end{equation}
Note that the differential Sharma-Mittal entropy of Gaussians may potentially be negative.

Figure~\ref{fig:SM} displays the plot of the Sharma-Mittal entropy for a $4\times 4$ covariance matrix set to $4I$, where $I$ denotes the identity matrix.

We also report respectively the R\'enyi, Tsallis and Shannon entropies for multivariate Gaussians  

\begin{eqnarray}
R_{\alpha}(\mu,\Sigma) &=& \log \frac{ (2\pi)^{\frac{d}{2}} |\Sigma|^{\frac{1}{2}}  }{\alpha^{\frac{d}{2(1-\alpha)}} },\label{eq:GauR}\\
T_{\alpha}(\mu,\Sigma) &=&  \frac{1}{1-\alpha} \left(
\frac{  \left( (2\pi)^{\frac{d}{2}}  |\Sigma|^{\frac{1}{2}} \right)^{1-\alpha} }{\alpha^{\frac{d}{2}}} -1
\right),\label{eq:GauT}\\
H(\mu,\Sigma) &=& \log \sqrt{(2\pi e)^d |\Sigma|} = \frac{1}{2} (d+ d\log 2\pi + \log |\Sigma| ).
\end{eqnarray}

\begin{figure}
\centering
\begin{tabular}{cc}
\includegraphics[width=0.5\textwidth]{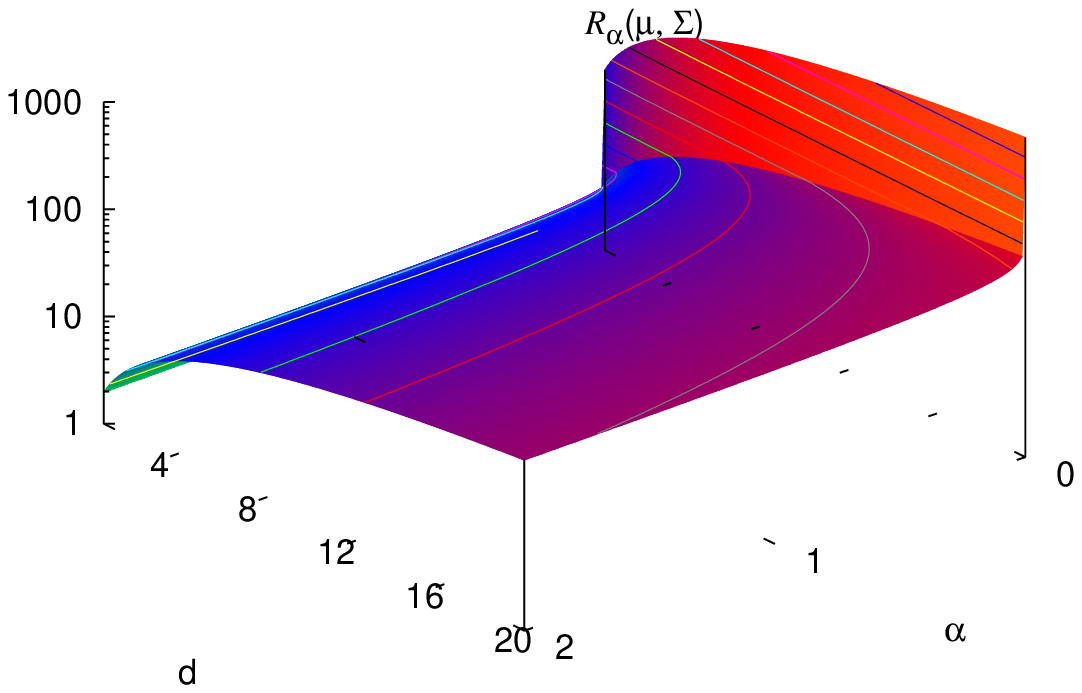} &
\includegraphics[width=0.5\textwidth]{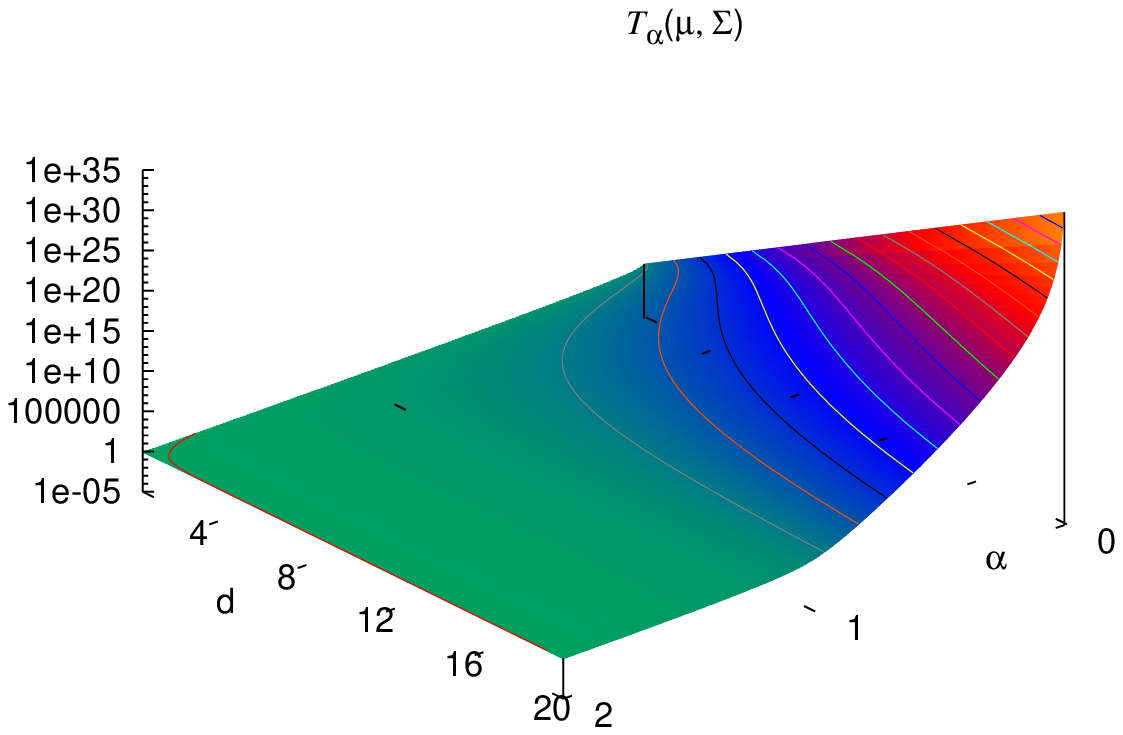} 
\\
(a) & (b) 
\end{tabular}

\caption{Plot of the (a) R\'enyi (Eq.~\ref{eq:GauR}) and (b) Tsallis (Eq.~\ref{eq:GauT}) entropies for  covariance matrices set to four times the matrix identity.
\label{fig:RT}}
\end{figure}

Figure~\ref{fig:RT} displays the plots of the R\'enyi (Eq.~\ref{eq:GauR}) and Tsallis   (Eq.~\ref{eq:GauT}) entropies  for a $d\times d$-dimensional covariance matrix $\Sigma=\sigma^2 I= 4I$ ($\sigma=2$).

Information geometry~\cite{informationgeometry-2000} considers the underlying differential geometry induced by a divergence.
From the Sharma-Mittal entropy, we can derive the Sharma-Mittal divergence~\cite{SharmaMittal-2005} between two distributions $P\sim p$ and $Q\sim q$  

\begin{equation}\label{eq:smdef}
D_{\alpha,\beta}(p : q) 
= 
\frac{1}{\beta-1} \left(\left(\int p(x)^{\alpha} q(x)^{1-\alpha} \dx\right)^{\frac{1-\beta}{1-\alpha}} -1 \right), \forall \alpha>0,\alpha\not =1,\beta\not=1.
\end{equation}
Note that $D_{\alpha,\beta}(p : q)=0$ if and only if $p=q$, since in that case $\int p(x)^{\alpha} q(x)^{1-\alpha} \dx=\int p(x)\dx=1$.

For $\alpha,\beta\rightarrow 1$, the divergence tends to the renown Kullback-Leibler divergence.
Let $C_\alpha(p:q)=\int p(x)^{\alpha} q(x)^{1-\alpha} \dx$ denote the $\alpha$-divergence~\cite{informationgeometry-2000},
related to the Hellinger integral of order $\alpha$: $H_\alpha(p,q)=1-C_\alpha(p,q)$.
For $\alpha=\frac{1}{2}$, the similarity measure $C_\frac{1}{2}(p:q)$ is symmetric and called the Bhattacharrya coefficient.
The Bhattacharrya coefficient is related to the
following squared Hellinger distance:

\begin{equation}
H^2(p:q)=\frac{1}{2}\int \left(\sqrt{p(x)}-\sqrt{q(x)}\right)^2 \dx = 1-C_{\frac{1}{2}}(p:q).
\end{equation}

We rewrite compactly the Sharma-Mittal divergence of Eq.~\ref{eq:smdef} as

\begin{equation}
D_{\alpha,\beta}(p : q) 
= 
\frac{1}{\beta-1} \left(C_\alpha(p,q)^{\frac{1-\beta}{1-\alpha}}-1  \right).
\end{equation}

Let us prove that for members $p(x)=p_F(x | \theta)$ and $q=p_F(x | \theta')$ belonging to the {\it same} exponential family $\E_F$, we have $C_\alpha(p:q)=e^{-J_{F,\alpha}(\theta:\theta')}$,
where 
\begin{equation}
J_{F,\alpha}(\theta:\theta') = \alpha F(\theta) + (1-\alpha) F(\theta') - F(\alpha\theta+(1-\alpha)\theta')
\end{equation}
is a Jensen difference divergence~\cite{BurbeaRao-1982}.

\noindent Proof:
\begin{eqnarray}
C_{\alpha}(p:q) & =&  \int p_F(x|\theta)^\alpha p_F(x|\theta')^{1-\alpha} \dx, \\
C_{\alpha}(\theta:\theta') & =& \int \exp^{\alpha(\inner{t(x)}{\theta}-F(\theta)+k(x))}\nonumber\\
&& \times 
\exp^{(1-\alpha)(\inner{t(x)}{\theta '}-F(\theta')+k(x))} \dx,\\
&=&   \int e^{ \inner{t(x)}{\alpha\theta+(1-\alpha)\theta'}} \nonumber\\
&&\times \exp^{-\alpha F(\theta)-(1-\alpha)F(\theta')+k(x)  }\dx,  \\
&=&  \int e^{F(\alpha\theta+(1-\alpha)\theta')-\alpha F(\theta)-(1-\alpha)F(\theta')} p_F(x;\alpha\theta+(1-\alpha)\theta')\dx, \label{eq:obs2} \\
&=&  e^{-J_{F,\alpha}(\theta:\theta')}  \int p_F(x;\alpha\theta+(1-\alpha)\theta')\dx, \\
&=& e^{-J_{F,\alpha}(\theta:\theta')}  >0.
\end{eqnarray}
Observe that for $\alpha\in(0,1)$, $\alpha\theta+(1-\alpha)\theta' \in\Theta$ since $\Theta$ is an open convex set, and therefore the distribution $p_F(x;\alpha\theta+(1-\alpha)\theta')$ is well-defined in Eq.~\ref{eq:obs2}.

It follows that the Sharma-Mittal divergence of distributions belonging to the same exponential family (even when $k(x)\not =0$) is the following closed-form formula

\begin{eqnarray} 
D_{\alpha,\beta}(p : q) &=& \frac{1}{\beta-1} \left( C_\alpha(\theta_p:\theta_q)^\frac{1-\beta}{1-\alpha}-1\right),\\
 &=& \frac{1}{\beta-1}   \left( e^{-\frac{1-\beta}{1-\alpha} J_{F,\alpha}(\theta_p:\theta_q)  } -1 \right).    \label{eq:SMdiv}
\end{eqnarray}

For multivariate Gaussians, let us explicit the Jensen difference divergence $J_{F,\alpha}$ as the difference of two terms 
using the $(\mu,\Sigma)$ coordinate system

\begin{eqnarray}
\alpha F(\theta) + (1-\alpha) F(\theta') 
&=& \frac{d}{2}\log 2\pi + \frac{1}{2}\log |\Sigma|^{\alpha} |\Sigma'|^{1-\alpha}\nonumber\\
&& +\frac{\alpha}{2}\mu^T\Sigma^{-1}\mu + \frac{1-\alpha}{2}{\mu'}^T {\Sigma'}^{-1}\mu'
\end{eqnarray}
and $F(\alpha\theta +(1-\alpha)\theta')$.
Let 
\begin{equation}
\bar{\theta}_\alpha=\alpha\theta +(1-\alpha)\theta'=(\alpha\Sigma^{-1}\mu+(1-\alpha){\Sigma'}^{-1}{\mu'},-\frac{\alpha}{2}\Sigma^{-1}-\frac{1-\alpha}{2}{\Sigma'}^{-1})=(\bar{v}_\alpha,\bar{M}_\alpha).
\end{equation}
Denote by 
$\bar{\Sigma}_\alpha=-\frac{1}{2}{\bar{M}_\alpha}^{-1}$ 
and 
$\bar{\mu}_\alpha=\bar{v}_\alpha\bar{\Sigma}_\alpha$ the corresponding parameters.
Using Eq.~\ref{eq:can}, we have  

\begin{equation}
F(\alpha\theta +(1-\alpha)\theta')=F(\bar{\mu}_\alpha,\bar{\Sigma}_\alpha)= \frac{1}{2}\log (2\pi)^d|\bar{\Sigma}_\alpha| + \frac{1}{2}{\bar\mu}_\alpha^T {\bar\Sigma}^{-1}_\alpha {\bar\mu}_\alpha.
\end{equation}

It follows that the Jensen difference divergence  between two Gaussian distributions $p\sim N(\mu,\Sigma)$ and $q\sim N(\mu',\Sigma')$ is given by the closed-form formula

\begin{equation}  \label{eq:SMdivGaussian}
J_{F,\alpha}(p : q) = \frac{1}{2} \left(
\log \frac{|\Sigma|^{\alpha} |\Sigma'|^{1-\alpha}}{|\bar{\Sigma}_\alpha|} 
+ \alpha\mu^T\Sigma^{-1}\mu  + (1-\alpha) {\mu'}^T {\Sigma'}^{-1}\mu' - {\bar\mu}_\alpha^T {\bar\Sigma}^{-1}_\alpha {\bar\mu}_\alpha 
\right),
\end{equation}
with
\begin{eqnarray}
\bar{\Sigma}_\alpha &=& \left(\alpha \Sigma^{-1} + (1-\alpha) {\Sigma'}^{-1} \right)^{-1},\\
\bar{\mu}_\alpha & = & \bar{\Sigma}_\alpha\bar{v}_\alpha= \bar{\Sigma}  \left(\alpha \Sigma^{-1}\mu+(1-\alpha){\Sigma'}^{-1}\mu'\right)
\end{eqnarray}

Letting $\Delta\mu=\mu'-\mu$, Eq.~\ref{eq:SMdivGaussian} can further be rewritten compactly as

\begin{equation}  \label{eq:SMdivGaussian2}
J_{F,\alpha}(p : q) = \frac{1}{2} \left(
\log \frac{|\Sigma|^{\alpha} |\Sigma'|^{1-\alpha}}{|\bar{\Sigma}_\alpha|} 
+ \alpha(1-\alpha)\Delta\mu^T  {\bar\Sigma}^{-1}_\alpha  \Delta\mu
\right).
\end{equation}

Thus we obtain a closed-form formula for the Sharma-Mittal divergence of multivariate Gaussians generalizing the R\'enyi $\alpha$-divergences, formerly reported in~\cite{EntropicGraphs-2002}:

\begin{eqnarray}
\lefteqn{D_{\alpha,\beta}( N(\mu,\Sigma) : N(\mu',\Sigma') )  =}\nonumber \\
&& \frac{ 1}{\beta-1}  \left ( e^{-\frac{1-\beta}{2(1-\alpha)} \left(
\log \frac{|\Sigma|^{\alpha} |\Sigma'|^{1-\alpha}}{|\bar{\Sigma}_\alpha|} 
+ \alpha(1-\alpha)\Delta\mu^T  {\bar\Sigma}^{-1}_\alpha  \Delta\mu
\right) } -1\right), \\
&& =  \frac{ 1}{\beta-1} \left( \left( \frac{|\Sigma|^{\alpha} |\Sigma'|^{1-\alpha}}{|\bar{\Sigma}_\alpha|}  \right)^{-\frac{1-\beta}{2(1-\alpha)}}
\exp \left({-\frac{\alpha(1-\beta)}{2} \Delta\mu^T  {\bar\Sigma}^{-1}_\alpha  \Delta\mu}\right) -1 \right). \label{eq:SMG}
\end{eqnarray}

\begin{figure}
\centering
\includegraphics[width=0.65\textwidth]{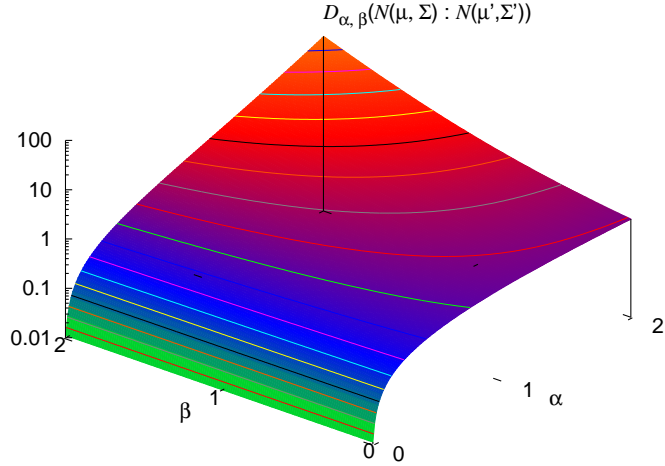} 

\caption{Plot of the Sharma-Mittal divergence $D_{\alpha,\beta}$ (Eq.~\ref{eq:SMG}) for univariate normal distributions with respective standard deviation $\sigma=\sqrt{2}$ and $\sigma'=2$, and mean difference $\mu-\mu'=4$.
\label{fig:SMdiv}}
\end{figure}

Figure~\ref{fig:SMdiv} shows the plot of the Sharma-Mittal divergence for univariate normal distributions with respective standard deviation $\sigma=\sqrt{2}$ and $\sigma'=2$, and mean difference $\mu-\mu'=4$ in Eq.~\ref{eq:SMG}. 
In practice, for numerical stability, we prefer to compute  the divergence by first computing the Jensen difference divergence of Eq.~\ref{eq:SMdivGaussian2}, and then applying generic formula of Eq.~\ref{eq:SMdiv}.

The underlying distribution is usually not explicitly given so that we need to first estimate the distribution or related quantities like its entropy~\cite{alphadivEstimation:2011}. 
Leonenko et al.~\cite{RenyiEstimation:2008} proposed a method to estimate entropies using the $k$-nearest neighbor graph ($k$-NN) of an independently and identically distributed sample set $x_1, ..., x_n$. However, their method suffers from the curse of dimensionality of computing $k$-NN graphs and falls short when dealing with moderate dimensions.
For exponential families, we can estimate the natural parameter of an exponential family using the maximum likelihood estimator (MLE) that admits the unique global optimum~\cite{Brown:1986,Naudts-2011} $\hat\theta$ such that

\begin{equation}\label{eq:MLE}
\nabla F(\hat\theta)=\frac{1}{n} \sum_{i=1}^n t(x_i).
\end{equation}

For multivariate Gaussians, from  the sufficient statistic $t(x)=(x, x^T x)$ we deduce that $\nabla F(\hat\theta)=(\hat\mu,\hat\Sigma+\hat\mu {\hat\mu}^T)=(\frac{1}{n}\sum_{i=1}^n x_i, \frac{1}{n}\sum_{i=1}^n x_i^T x_i)$.

It follows a simple and fast scheme to estimate the Sharma-Mittal entropy (or divergence) from $n$ observations sampled identically and independently from an exponential family distribution: Estimate the natural parameter using Eq.~\ref{eq:MLE} and apply formula of Eq.~\ref{eq:SMEnok}.

To conclude, let us note that {\it any} arbitrary smooth density can be approximated by an exponential family of order depending on the approximation precision~\cite{Cobb:1983:EMR,UniversalEF-2007} (enforcing no extra auxiliary carrier measure: That is, with $k(x)=0$).
Thus we can approximate the Sharma-Mittal entropy of an arbitrary probability density~\cite{UniversalEF-2007} by approximating it to a close exponential family, and then applying the closed-form formula Eq.~\ref{eq:SMEnok}. 
We believe that Eq.~\ref{eq:SMEnok}, Eq.~\ref{eq:entGau}, Eq.~\ref{eq:SMdiv} (numerically stable) and~\ref{eq:SMG} will prove useful when experimenting for suitable parameters $(\alpha,\beta)$ in various statistical signal tasks~\cite{EntropicGraphs-2002}.

\section*{Acknowledgments}
The authors would like to thank the anonymous reviewers for their helpful comments and suggestions.
Frank Nielsen (5793b870) would like to thank Dr. Kitano and Dr. Tokoro for their  support.
Additional material is available on-line at \url{http://www.informationgeometry.org/SharmaMittal/}

%%%%%%%%
\section*{References}
%%%%%%%%

\end{document}